\documentclass{emulateapj}
\usepackage{apjfonts}
\usepackage{epsfig}
\usepackage{natbib}

\newcommand{\lcdm}{$\Lambda$CDM}
\newcommand{\tten}{\times 10}
\newcommand{\gas}{\mathrm{gas}}
\newcommand{\Msun}{M_{\sun}}

\shorttitle{Disk Galaxy Formation in a \lcdm\ Universe}
\shortauthors{Robertson et al.}

\begin{document}

\title{Disk Galaxy Formation in a \lcdm\ Universe}
\author{Brant Robertson\altaffilmark{1,4},
        Naoki Yoshida\altaffilmark{2}, 
        Volker Springel\altaffilmark{3}, and
        Lars Hernquist\altaffilmark{1}} 

\altaffiltext{1}{Harvard-Smithsonian Center for Astrophysics, 
        60 Garden St., Cambridge, MA 02138, USA}
\altaffiltext{2}{National Astronomical Observatory Japan, Osawa 2-21-1, Mitaka, Tokyo 181-8588, Japan}
\altaffiltext{3}{Max-Planck-Institut f\"ur Astrophysik, Karl-Schwarzschild-Stra\ss e 1, 85740 Garching bei M\"unchen, Germany}
\altaffiltext{4}{brobertson@cfa.harvard.edu}

\begin{abstract}

We describe hydrodynamical simulations of galaxy formation in a
$\Lambda$ cold dark matter (CDM) cosmology performed using a
subresolution model for star formation and feedback in a multiphase
interstellar medium (ISM). In particular, we demonstrate the formation
of a well-resolved disk galaxy.  The surface brightness profile of the
galaxy is exponential, with a $B$-band central surface brightness of
21.0 mag arcsec$^{-2}$ and a scale-length of $R_{\rm d} = 2.0\,
h^{-1}{\rm kpc}$.  We find no evidence for a significant bulge
component.  The simulated galaxy falls within the $I$-band
Tully-Fisher relation, with an absolute magnitude of $I$ = --21.2 and
a peak stellar rotation velocity of $V_{\rm rot}=121.3 \,{\rm km\,
s^{-1}}$.  While the total specific angular momentum of the stars in
the galaxy agrees with observations, the angular momentum in the inner
regions appears to be low by a factor of $\sim 2$.  The star formation
rate of the galaxy peaks at $\sim 7$ M$_{\sun}$ yr$^{-1}$ between
redshifts $z=2-4$, with the mean stellar age decreasing from $\sim 10$
Gyrs in the outer regions of the disk to $\sim 7.5$ Gyrs in the
center, indicating that the disk did not simply form inside-out.  The
stars exhibit a metallicity gradient from $0.7$ Z$_{\sun}$ at the edge
of the disk to $1.3$ Z$_{\sun}$ in the center.  Using a suite of
idealized galaxy formation simulations with different models for the
ISM, we show that the effective pressure support provided by star
formation and feedback in our multiphase model is instrumental in
allowing the formation of large, stable disk galaxies.  If ISM gas is
instead modeled with an isothermal equation of state, or if star
formation is suppressed entirely, growing gaseous disks quickly
violate the Toomre stability criterion and undergo catastrophic
fragmentation.

\end{abstract}


\section{Introduction}
\label{sec:intro}

Numerical simulations have become one of the most important
theoretical tools for exploring the complicated problem of galaxy
formation.  Modern numerical work on galaxy formation was initiated by
\cite{kg91a}, \cite{nb91a}, and \cite{katz92a}, who included radiative
cooling by hydrogen and helium, and attempted to account for star
formation and feedback processes.  Subsequent work employed more
sophisticated treatments of supernova feedback \citep{nw93a}, models
for chemical enrichment \citep{sm94a}, and methods to `zoom in' on
galaxies of interest within cosmological simulations \citep{nw94a}

While these studies achieved limited success in creating rotationally
supported disks, the simulated galaxies typically failed to reproduce
their observed counterparts.  In general, the simulated disks were
found to be too small and were more centrally concentrated than actual
galaxies \citep{nw93a, nw94a, nfw95a}.  In addition, the star
formation in the models was overly efficient, converting too large a
fraction of the gas into stars by the present day
\citep{nfw95a,sm94a,sm95a}.

Many of the shortcomings of the early modeling can be tied to the
coarse resolution of the simulations and the manner in which feedback
was treated.  In a cosmological context, the multiphase structure of
the ISM within galaxies cannot be resolved in detail, so star-forming
gas is generally described as a single-phase medium.  If feedback
energy is deposited into this gas purely as thermal energy, it will be
quickly radiated away, and any impact of feedback from star formation
will be lost.  In addition, if star formation is very efficient in low
mass halos at high redshifts, a large number of dense, compact
galaxies will form and eventually be incorporated into the inner
regions of larger galaxies through hierarchical merging.  If these
baryonic clumps lose angular momentum to dark halos, the resulting
objects will be smaller in extent than if the baryons had been
accreted smoothly.  Thus, an early collapse of baryons opens a channel
for gas to lose angular momentum, yielding galaxies that are too
compact, lack angular momentum, and produce stars overly efficiently
compared with observations.

Subsequent work attempted to alleviate these shortcomings by employing
stronger forms of feedback and modifications to the cosmological
framework.  In some cases, the problems noted above were exacerbated.
For example, photoheating by a diffuse ultraviolet (UV) background was
found to further reduce the angular momentum content of the simulated
galaxies \citep{ns97a}.  Other efforts included the impact of
preheating and gas blow-out from small halos \citep{slgv99a} and the
formation of galaxies in a warm dark matter (WDM) cosmology
\citep{sld01a}.  More direct comparisons to observations were made
possible by incorporating spectral synthesis techniques into the
modeling \citep{csfv98a} and by using the Tully-Fisher relation to
constrain the hierarchical origin of galaxies \citep{sn99a, ns00a}.
However, the galaxies in these simulations were still too
concentrated.

The most recent studies of disk formation have yielded somewhat more
promising results.  Using a model of self-propagating star formation
combined with supernova feedback and a UV background,
\cite{slgp02a,slgp03a} produced disk galaxies deficient in angular
momentum by less than an order of magnitude.  \cite{governato02a}
report the formation of realistic disk galaxies in $\Lambda$CDM and
$\Lambda$WDM cosmologies using simulations that include standard
prescriptions for cooling, a UV background, and star formation.
\cite{abadi03a, abadi03b} present detailed analyses of simulated
galaxies with kinematic and photometric properties similar to observed
Sab galaxies.

Although the various attempts to simulate disk formation have provided
some impressive successes, the essential physics behind this process
remains unclear.  \cite{governato02a} claim that the low angular
momentum of simulated galaxies in previous works owed at least partly
to inadequate resolution.  In addition, they emphasize the impact that
the matter power spectrum can have on galaxy formation through
simulations of WDM universes in which the bulge and spheroid
components of galaxies are smaller than in CDM models.  These findings
are supported by \citet{slgp02a,slgp03a} and \citet{sld01a}, whose
calculations indicate that high resolution, strong stellar feedback,
and warm dark matter can produce realistic disk galaxies.  Finally,
\cite{abadi03a} suggest that a crucial ingredient for forming proper
disks is an implementation of feedback that can regulate star
formation.

Here, we use a `multiphase model' to describe star-forming gas, in
which the ISM consists of cold clouds in pressure equilibrium with an
ambient hot phase \citep{sh03a}.  Radiative cooling of gas leads to
the growth of clouds, which in turn host the material that fuels star
formation. The supernovae associated with star formation provide
feedback by heating the ambient medium and evaporating cold
clouds. The feedback treatment establishes a self-regulated cycle for
star formation, pressurizing the star-forming gas.  Implemented in a
subresolution manner, our approach makes it possible to obtain
numerically converged results for the star formation rate at moderate
resolution.  This feature is particularly important for simulations of
CDM cosmologies, where early generations of galaxies may be difficult
to resolve.

Using this model of star-forming gas, \cite{sh03b} obtained a
converged prediction for the history of cosmic star formation that
agrees well with observations at redshifts $z \le 4$ \citep{sh03b,
her03}.\footnote[1]{We note an error in figure 12 of \citet{sh03b}
where the observational estimates of the star formation rate are
plotted too high by a factor of $h^{-1} = 1.4$.  When corrected, the
observed points are in better agreement with the theoretical
estimates; see astro-ph/0206395 and \citet{nachos03}.}  In subsequent
work, \cite{nsh03a, nsh03b, nshm03} showed that this model also
accounts for the observed abundance and star formation rates of damped
Lyman-alpha absorbers and Lyman-break galaxies at $z\sim 3$.

In what follows, we study the consequences of our multiphase model for
the ISM on the formation of disk galaxies.  First, we employ a
high-resolution simulation to identify realistic
disk galaxies in a cosmological context.  For one such disk galaxy, we
examine its kinematic and photometric properties in detail,
demonstrating good agreement with observations of local spirals.
Second, we use a set of idealized simulations to study disk formation
in individual dark matter halos to isolate the effect of different
models for the equation of state of the ISM.  This analysis
demonstrates that the feedback in our multiphase model alters the
dynamics by pressurizing the star-forming gas and stabilizing forming
disks against fragmentation.  We emphasize that this aspect of our
modeling does not depend on the details of our prescription for star
formation and feedback, but is determined by the effective equation of
state for star-forming gas.  Thus, our conclusions should obtain
generally, provided that the actual bulk equation of state for the ISM
has characteristics similar to those of our description.

In \S~\ref{sec:sim:cosmo}, we present our simulation method. We review
our analysis procedure in \S~\ref{sec:analysis}, and discuss our
findings for the structural (\S~\ref{sec:disk:nonkin}) and kinematic
properties (\S~\ref{sec:disk:kin}) of one simulated disk galaxy.  In
\S~\ref{sec:sim:halo:intro}, we describe our idealized simulations and
their results.  Finally, we conclude and suggest directions for
further research in \S~\ref{sec:conclusions}.

\section{Cosmological Simulation}
\label{sec:sim:cosmo}

Our simulations were performed using the parallel $N$-body/smoothed
particle hydrodynamics (SPH) code {\small GADGET2} in its
``conservative entropy'' formulation \citep{sh02a}, to mitigate
problems with lack of energy and entropy conservation in older
treatments of SPH \citep[e.g.][]{her93,oshea03}.  We adopt a flat
$\Lambda$CDM cosmology with cosmological parameters $\Omega_{\rm
m}=0.3$, $\Omega_{\Lambda}=0.7$, $\Omega_{\rm b}=0.04$, and
$\sigma_{8}=0.9$, and set the primordial power spectrum index to
$n=1$.  Throughout, we select a value for the Hubble constant of
$H_{0} = 100\,h\,{\rm km\, s^{-1} Mpc^{-1}}$ with $h=0.7$.

For our cosmological simulation, we populate a periodic volume of 10
$h^{-1}$Mpc on a side with $144^{3}$ low-resolution dark matter (LRDM)
particles.  At the center of the box, a 5 $h^{-1}$Mpc cubic region is
selected as a high resolution region where we replace the LRDM
particles with particles of eight-times lower mass.  The initial
displacement field is then calculated following a standard ``zooming''
procedure \citep{tormen97a, power03a}, where small scale perturbations
are added appropriately in the high-resolution region.  Note that the
high-resolution zone does not target a particular object, with the
intent of eliminating bias that could be introduced by selecting halos
that may be intrinsically favorable for disk galaxy formation.  We
further split the high-resolution particles into dark matter (HRDM)
and gas.  The resulting particle masses of each component are then
$m_{\rm LRDM}=2.79\times 10^{7} h^{-1} M_{\odot}$, $m_{\rm
HRDM}=3.02\times 10^{6} h^{-1} M_{\odot}$, and $m_{\rm gas}=4.65\times
10^{5} h^{-1} M_{\odot}$.  We set the gravitational softening length
for the high-resolution particles to 0.65 comoving $h^{-1}$kpc.  While
our 10 $h^{-1}$ Mpc box is too small to be fully representative of the
$z=0$ universe, the volume is sufficient for our purposes as our
current work does not concern, for example, large-scale correlations
of galaxies or the global mass function.  Here, we are interested only
in individual, galactic-sized objects which are mostly unaffected by
the simulation box size.

We include a UV background and radiative cooling and heating in the
manner of \cite{kwh96} and \cite{dave99}, as well as star formation,
supernova feedback, and metal enrichment.  We employ the multiphase
model developed by \cite{sh03a} to describe the star-forming gas
\citep[see also][]{yepes97,hp99}.  Our approach accounts for some of
the key aspects of the multiphase structure of the ISM \citep{mo77a}
without spatially resolving the different phases explicitly. Instead,
a statistical mixture of the phases is computed analytically, taking
into account the growth of cold clouds embedded in a supernova-heated
ambient phase, the formation of stars out of the cloud material, and
the evaporation of clouds in supernova remnants.

Our strategy is motivated by two basic limitations in modeling star
formation on cosmological scales.  First, there is no fundamental
theory of this process.  Second, large-scale simulations lack the
resolution that would be required to characterize the ISM in detail.
Our statistical method decouples our ignorance of star formation from
its impact on galactic scales, by coarse-graining the physics of
star-forming gas.  In principle, this strategy could be applied to any
model of star formation.

\begin{figure}[t]
\centering
\epsfig{file=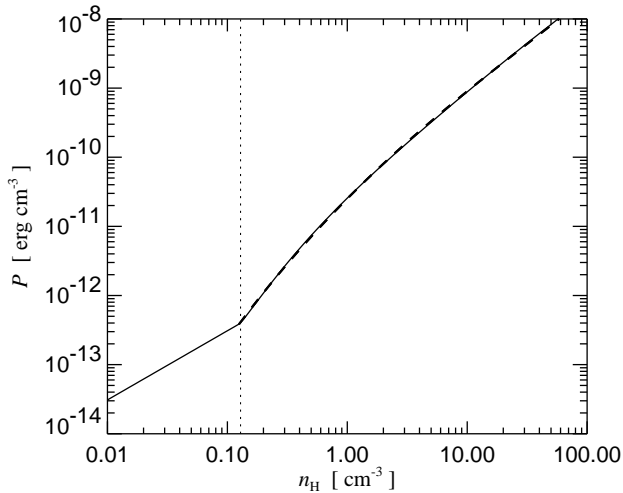,angle=0,totalheight=2.75in,keepaspectratio=true}
\vskip -1mm
\figcaption[f1.eps]{\label{fig:eos}
Effective equation of state for star-forming gas in our multiphase
model \citep{sh03a}.  The effective pressure, $P_{\rm eff}$, is
plotted in cgs units versus the total gas density in hydrogen atoms
per cubic centimeter.  The solid line is the exact equation of state,
while the thick dashed line is a simple fit (see text), which is
accurate to about $1\%$.  The vertical dotted line shows the
transition between an isothermal gas (to the left) and the effective
equation of state in our multiphase model (to the right).}
\vskip -3mm
\end{figure}

For a detailed description of our multiphase model we refer the reader
to \cite{sh03a}. For definiteness, we list in Table 1
the parameter values adopted in our simulations.
The parameters include the efficiency of the evaporation process
$A_{0}$, the mass fraction of stars that are short-lived and die as
supernovae $\beta$, the cold gas cloud temperature $T_{\rm cloud}$,
the effective supernova temperature $T_{\rm SN}$, and the gas
consumption time-scale $t_0^\star$.  As \cite{sh03a} argue, all of
these aside from one, e.g. $t_0^\star$, can be constrained by simple
physical arguments. We adjust $t_0^\star$ to match the Kennicutt Law
which describes empirically observed star formation rates in nearby
galaxies \citep[e.g.][]{kennicutt89a,kennicutt98a}.

On the surface, our description of star-forming gas appears
complicated, with the properties of the different phases described by
differential equations that include the various processes listed
above.  However, in the end, our model can be reduced to two main
ingredients that make our results easier to interpret.
\cite{sh03a} showed that the star-forming gas quickly establishes a
self-regulated cycle in which the mass and energy of the phases can be
approximated by equilibrium solutions.  In this limit, the impact of
star formation and feedback can be reduced to: 1) the rate at which
gas is converted into stars, and 2) the effective equation of state
for star-forming gas.

Here, as in \citet{sh03a}, we choose to parameterize the star
formation rate by
\begin{equation}
{{{\rm d}\rho_*}\over{{\rm d}t}} \, = \, (1-\beta) {{\rho_c}\over{t_*}} \, ,
\label{eq:sfr}
\end{equation}
where
\begin{equation}
t_* \, = \, t_0^\star \left ( {{\rho}\over{\rho_{\rm th}}}
\right ) ^{-1/2} \, .
\end{equation}
The free parameter $t_0^\star$ is chosen to match the Kennicutt Law,
$\rho_{\rm th}$ is a threshold density above which gas is subject to
thermal instability and star formation, and $\rho_c$ is the density of
clouds which, in equilibrium, is given by equation (18) of
\cite{sh03a}. \cite{sh03a} showed that $\rho_{\rm th}$ can be fixed by
e.g.~requiring the equation of state to be continuous.

\begin{figure}[t]
\centering
\epsfig{file=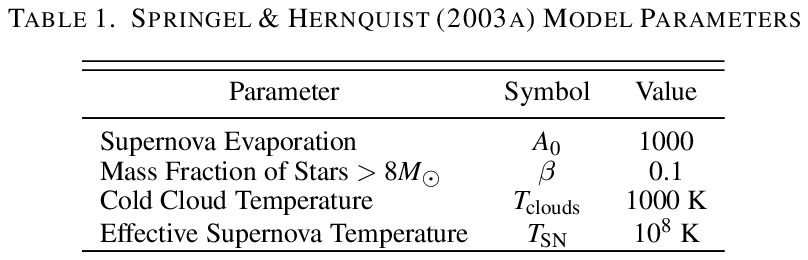,angle=0,keepaspectratio=true}
\end{figure}

In our subresolution model, feedback from supernovae adds thermal
energy to the ISM, pressurizing it and modifying its equation of
state.  On the scales relevant to the dynamics of the gas, this is
described by an effective equation of state, which can be written
\begin{equation}
P_{\rm eff} \, = \, (\gamma - 1) \, \rho \, u_{\rm eff} ,
\end{equation}
where $\gamma$ is the ratio of specific heats, $\rho$ is the total gas
density, and $u_{\rm eff}$ is the effective specific thermal energy of
the star-forming gas which, in equilibrium, is given by equation (19)
of \cite{sh03a}.  The effective equation of state for the particular
model parameters employed in our study is illustrated in Figure
\ref{fig:eos}, in cgs units.  Also shown in Figure \ref{fig:eos} is a
simple fit to the equation of state, which is accurate to about $1\%$
and is given by
\begin{eqnarray}
\log P_{\rm eff} \, &=& \, 0.050 \, (\log n_{\rm H})^3 \, - \, 
0.246 \, (\log n_{\rm H})^2 \, + \\
& &\, 1.749 \, \log n_{\rm H} \, - \,
10.6 \, , \,\,\,
{\rm for} \,\, \log n_{\rm H} > -0.89 \, . \nonumber
\label{eq:eosfit}
\end{eqnarray}
In the particular example shown in Figure \ref{fig:eos}, the gas is
assumed to be isothermal for densities $\rho \le \rho_{\rm th}$.  In
cgs units, for the parameter choices adopted here, the critical
density at which the gas begins to depart from isothermality is
$n_{\rm H, th} = 0.128$ cm$^{-3}$.  This choice ensures that the
effective pressure is a continuous function of density.  For the
equation of state shown in Figure 1, a different value of $n_{\rm H,
th}$ would introduce an unphysical jump in pressure at the transition
density where the gas is no longer isothermal.

\begin{figure*}[t]
\centering
\epsfig{file=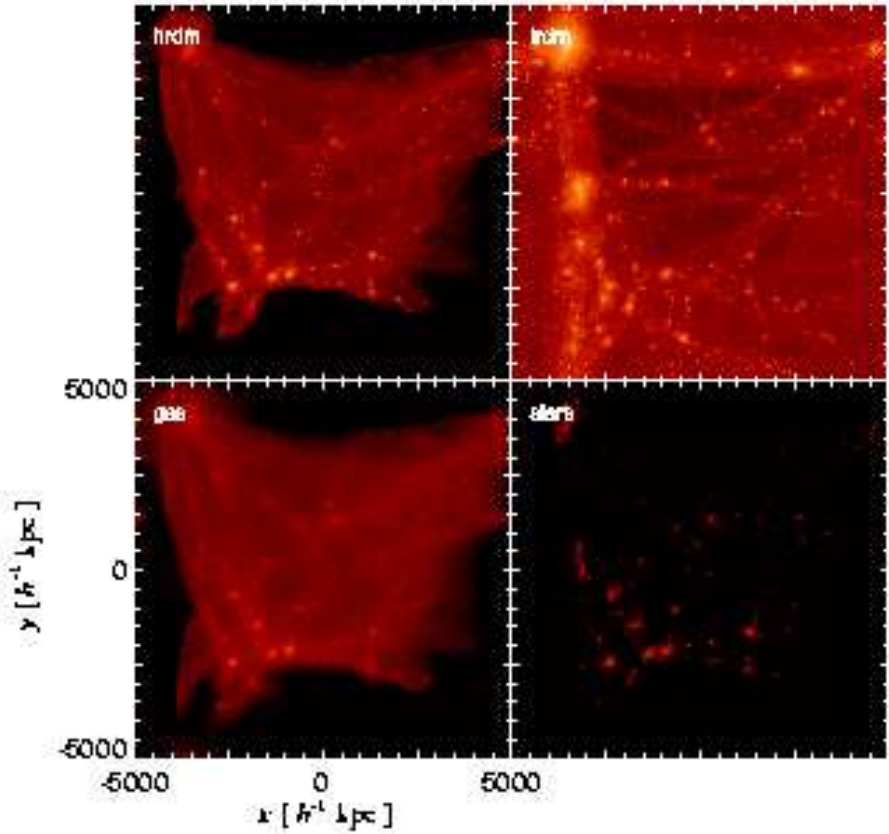,angle=0,totalheight=5.in,keepaspectratio=true}
\vskip -1mm
\figcaption[f2.eps]{
\label{fig:final}
Entire simulation volume at $z=0$ projected onto a 10 $h^{-1}$ Mpc
square area, showing the end-state of the simulation.  The upper left
and right panels show the HRDM and LRDM particles, respectively.  The
lower left and right panels show the gas and star particles,
respectively.  Owing to gravitational effects the low- and
high-resolution particles have intermingled at the boundary of the
high-resolution region, but the LRDM particles have not penetrated the
inner high-resolution region.  Star particles have spawned in regions
where the gas density has exceeded the critical density for star
formation.
}
\vskip -5mm
\end{figure*}

Our implementation of feedback differs significantly from previous
works \citep[e.g.][]{slgp03a,governato02a}.  In our subresolution
model, feedback energy is ``stored'' in the surrounding gas, adding
pressure support to it.  As shown in Figure \ref{fig:eos}, at
densities $\rho > \rho_{\rm th}$, the equation of state becomes
stiffer than isothermal due to this feedback energy.  This
pressurization regulates star formation occurring within the ISM and,
as we discuss later, modifies the bulk dynamics of the gas by altering
the pressure gradient term in Euler's equation.  It is this property
of our description that we believe is responsible for many of the
differences between our numerical results and earlier simulations.

In passing, we note that for the purposes of large-scale dynamics, our
multiphase model can be reduced to the choice of star formation rate,
e.g.~equation~(\ref{eq:sfr}), and the effective equation of state of
star-forming gas, e.g.~equation~(\ref{eq:eosfit}) and
Figure~\ref{fig:eos}.  The results presented here could be obtained
with any hydrodynamical code that uses the same expression for the
star formation rate and an equation of state like that in Figure
\ref{fig:eos}, without specific reference to all the complexities of
our multiphase model.

In Figure \ref{fig:final}, we show the evolved simulation volume at
$z=0$, in projection.  The upper and lower left panels show the
high-resolution dark matter (HRDM) and gas particles, respectively.
The high-resolution region of the simulation has increased in size
from the initial 5 $h^{-1}$ Mpc cubic region at the center of the
simulation, owing to the growth of gravitational structure.  Some of
the low- and high-resolution particles have mixed in the boundary
region of the low-resolution zone, whose shape has also been
distorted.  However, the LRDM particles (upper right panel) have not
penetrated into the inner regions of the high-resolution region where
many luminous galaxies have formed out of the gas, as marked by the
star particles (lower right panel).

\section{Disk Galaxy Selection}
\label{sec:analysis}

We use a standard friends-of-friends (FOF) group-finding algorithm to
identify dark matter halos, restricting ourselves to objects comprised
of high-resolution particles.  We then identify the most bound star or
gas particle within each object and produce a new collection of
particles for each halo that encompasses all particles that lie within
the radius $R_{200}$ that encloses 200 times the critical density
around these most-bound baryonic particles.  The group catalogue
produced from this procedure serves as our primary galaxy sample.

From our galaxy catalogue, we have visually inspected the 10 largest
objects, measured their angular momenta, and selected a large disk
galaxy for further detailed analysis in this paper.  This galaxy
displays the most disk-like morphology of all the galaxies in the
simulation resolved with a similar number of particles.  While we do
find other flattened systems in the simulation, the majority of them
do not display such well defined disk-like morphology and kinematics
as the galaxy described below.

\begin{figure*}[t]
\centering
\epsfig{file=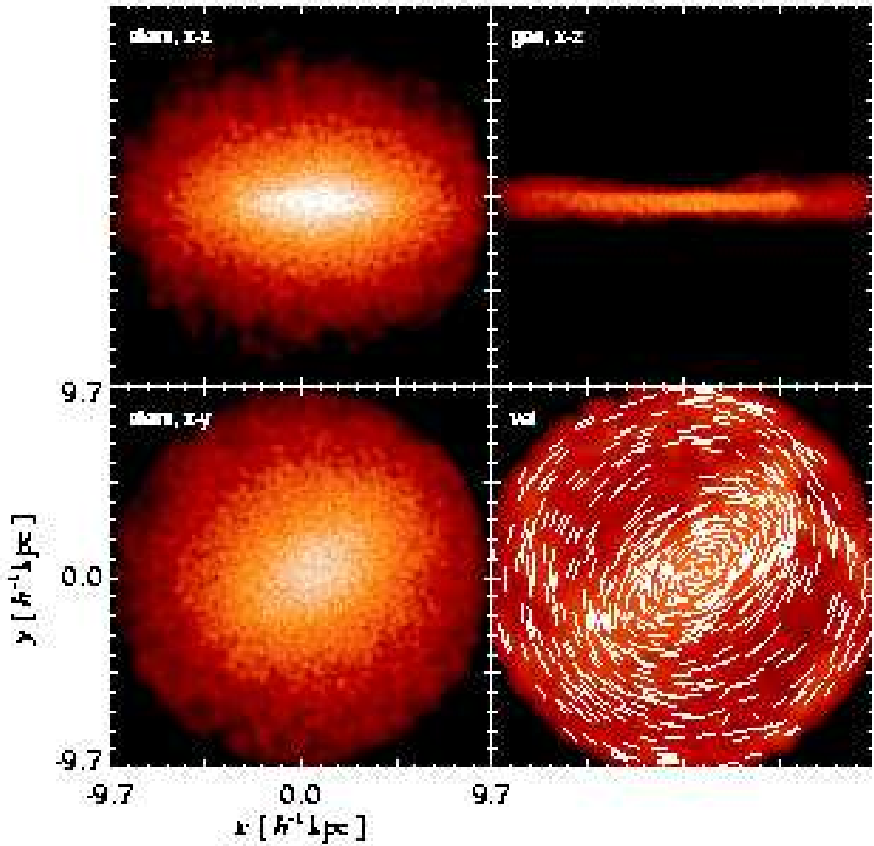,angle=0,totalheight=5.in,keepaspectratio=true}
\vskip -1mm
\figcaption[f3.eps]{
\label{fig:galaxy}
Disk galaxy at $z=0$, displaying the star (left panels) and gas (right
panels) particle distributions within $0.1\,R_{200}$, projected
perpendicular (upper panels) and parallel (lower panels) to the
direction of the stellar angular momentum vector.  The lower right
panel also traces the gas velocity field and demonstrates the
rotational support of the galaxy.
}
\vskip -5mm
\end{figure*}

The galaxy analyzed here, which we term ``galaxy C1'' for simplicity,
has $61,372$ dark matter, $21,506$ gas, and $88,138$ star particles
within $R_{200}$ at $z=0$. Note that the star particle mass was set to
half the original mass of the gas particles, so the overall baryon
fraction within $R_{200}$ is 14.1\%; slightly larger than the
universal baryon fraction of 13.3\%. Figure \ref{fig:galaxy} shows the
star (left panels) and gas (right panels) particle distributions
within $0.1\,R_{200}$ at $z=0$, projected perpendicular (upper panels)
and parallel (lower panels) to the direction of the stellar angular
momentum.  In addition, the lower right panel traces the gas velocity
field in the plane of the disk.  The gas in the galaxy has collapsed
into a rotationally supported disk, surrounded by a thicker stellar
disk.  The velocity field of the galaxy illustrates the mainly
circular trajectories of the gas particles in the disk plane.  A weak
bar is visible in the disk gas, but is not strong in the stars that
dominate the baryonic mass of the galaxy.  We note that the stellar
component is slightly elongated in the direction of the bar, however.
The galaxy has reached a mass of $M_{200}=2.16 \tten^{11} h^{-1}$
M$_{\sun}$ at $z=0$, with a stellar mass of $M_{\star} = 2.05
\tten^{10} h^{-1}$ M$_{\sun}$ and a gas mass of $M_{\gas} = 9.44
\tten^{9}\, h^{-1}\,{\rm M}_{\sun}$.  The baryonic mass of the galaxy
is then 68\% stellar and 32\% gaseous.

\begin{figure*}[t]
\centering
\epsfig{file=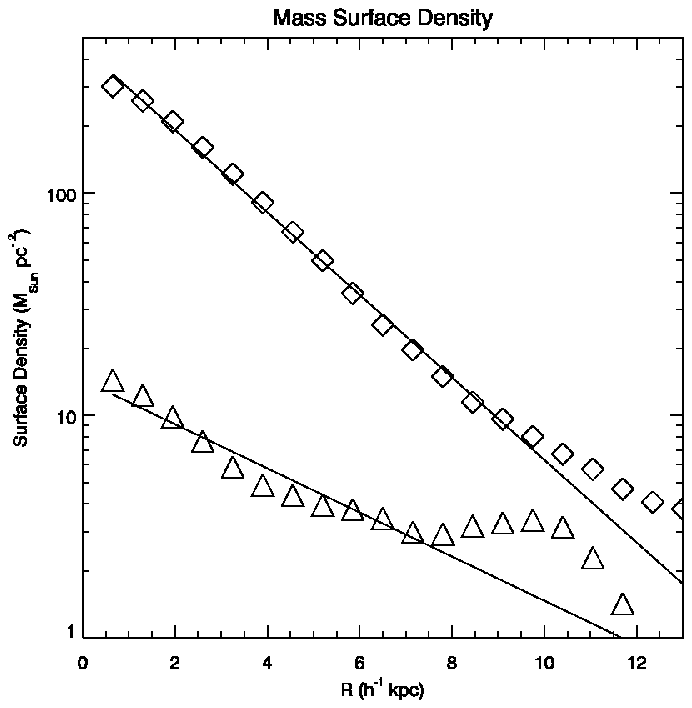,angle=0,totalheight=3.in,keepaspectratio=true}
\epsfig{file=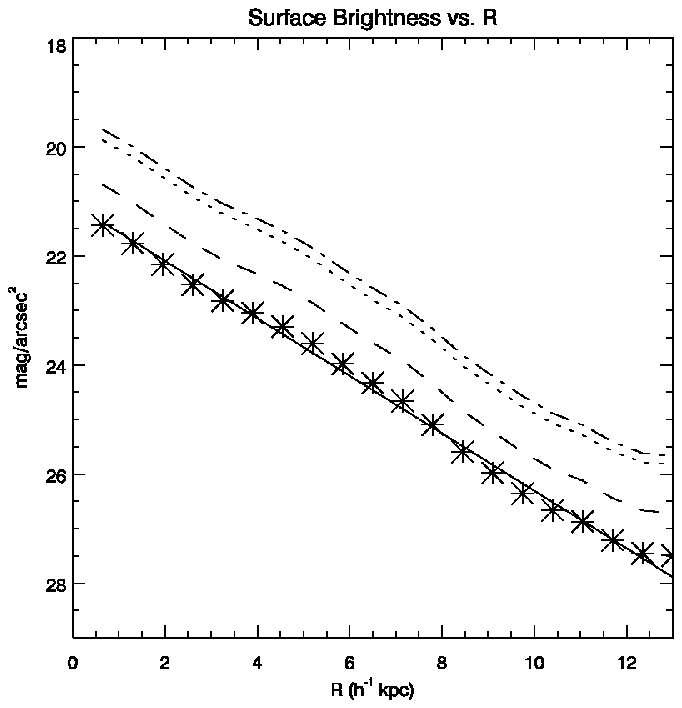,angle=0,totalheight=3.in,keepaspectratio=true}
\vskip -1mm
\figcaption[f4.eps]{
\label{fig:smd}
Mass surface density for the stars (diamonds) within $30\,h^{-1}$ kpc
of the galaxy center and gas particles (triangles) within $2\,h^{-1}$
kpc of the disk plane.  Using an exponential profile
$\Sigma(r)=\Sigma_{0}\exp(-r/R_{d})$ to fit each component separately,
we find best fit values of $\Sigma_{0,\star} = 452 h
\mathrm{\,\,M_{\sun}} \mathrm{pc}^{-2}$, $R_{d,\star} = 2.3 h^{-1}
\mathrm{kpc}$ for the stars and $\Sigma_{0,\gas} = 15.2 h
\mathrm{\,\,M_{\sun}} \mathrm{pc}^{-2}$, $R_{\gas} = 4.7 h^{-1}
\mathrm{kpc}$ for the gas.  The least squares fit for the stars (solid
line) and gas (dashed line) are plotted over the measured values.
}
\figcaption[f5.eps]{
\label{fig:sb}
Surface brightness profile of the galaxy in $B$ (stars), $V$ (dashed
line), $I$ (dashed-dotted line), and $K$ (dotted line) bands.  The
solid line is the fit to the B-band profile, with best fit central
surface brightness $\mu_{0,B}=21.0$ mag arcsec$^{-2}$ and B-band disk
scale length $R_{d,B}=2.0 h^{-1}$ kpc.  The standard central B-band
surface brightness value for disk galaxies is
$\mu_{B,\mathrm{Freeman}}=21.65$ mag arcsec$^{-2}$.  The disk
brightness is exponential in each band throughout the inner 10
$h^{-1}$ kpc of the galaxy.
}
\vskip -5mm
\end{figure*}

\section{Structural Properties of Galaxy C1}
\label{sec:disk:nonkin}

\subsection{Mass surface density profiles}
\label{subsec:smd}

Figure \ref{fig:smd} shows the stellar (diamonds) and gas (triangles)
mass surface densities of galaxy C1 projected along the stellar
angular momentum vector onto the disk plane.  The stellar profile
includes particles within $30\, h^{-1}$ kpc of the disk center, while
the gas profile includes particles only within $2\,h^{-1}$ kpc of the
disk plane.  The stellar and gas mass surface densities are roughly
exponential out to $r=8\, h^{-1}$ kpc.  If we fit them with an
exponential of the form
\begin{equation}
\Sigma(r) = \Sigma_{0}\,{\rm e}^{-r/R_{d}},
\end{equation}
\noindent
we find best fit values of $\Sigma_{\rm 0, \star} = 452\, h\,
\mathrm{\,\,M_{\sun}} \mathrm{pc}^{-2}$, $R_{\mathrm{d,\star}} = 2.3\,
h^{-1} \mathrm{kpc}$ for the stars and $\Sigma_{\rm 0,gas} = 15.3\, h
\mathrm{\,\,M_{\sun}} \mathrm{pc}^{-2}$, $R_{\mathrm{d},\gas} = 4.7\,
h^{-1} \mathrm{kpc}$ for the gas.  In the plot, we include least
squares fits to the stellar and gas mass surface densities.  For
comparison, the stellar disk of the Milky Way has a scale-length
$R_{\rm MW}\sim 3.5$ kpc, similar to the physical scale-length of
$\sim 3.3$ kpc for our simulated galaxy C1.  We note that the gas mass
surface density begins to diverge from the exponential fit between
$8-10\, h^{-1}$ kpc, at the edge of the gaseous disk.

\subsection{Photometric properties}
\label{subsec:phot}

From the known masses, metallicities, and ages of the star particles,
we use a population spectral synthesis code to calculate the SDSS
$ugriz$ \citep{fukugita96a} AB system magnitudes and $JHK$ magnitudes
for each particle.  The population synthesis code assumes a
\cite{kroupa01a} IMF and limits input metallicities to a range of
0.005 Z$_{\sun}$ $<$ Z $<$ 2.5 Z$_{\sun}$, outside of which we adopt
the minimum or maximum value.  We then use the \cite{fukugita96a}
conversions from SDSS colors to the Johnson-Morgan-Cousins $UBVRI$
\citep{jm53a, cousins78a} Vega system to calculate the absolute
magnitude of the star particles in each band.  We bin star particles
within $30\, h^{-1}$ kpc of the center of the galaxy into annuli and
calculate their total luminosities to determine a surface brightness
profile.  The measured surface brightness of galaxy C1 in $B$, $V$,
$I$, and $K$-bands is plotted in Figure \ref{fig:sb}.  The best fit
$B$-band central surface brightness is $\mu_{0,B}=21.0$ mag
arcsec$^{-2}$ with a $B$-band luminosity scale-length of $R_{d,B} =
2.0$ $h^{-1}$ kpc.  Our measured value for the $B$-band central
surface brightness agrees well with the canonical value of
$\mu_{0,B}=21.65$ mag arcsec$^{-2}$ for spiral galaxies
\citep{freeman70a}.

Figure \ref{fig:sb} demonstrates that the surface brightness profile
of galaxy C1 is exponential from the very center of the galaxy out to
$r \sim 10\, h^{-1}$ kpc, and
shows no evidence for a bulge component.  To our knowledge, this
object is the first published example of a galaxy formed within a
cosmological simulation that displays an exponential surface
brightness profile with no significant bulge.

For reference, the $u$, $g$, $v$, $r$, $i$, $z$, $U$, $B$, $R$, $I$,
$J$, $H$, and $K$-band absolute magnitudes of the galaxy, measured by
summing the luminosities of all stars used to calculate the surface
brightness profiles out to $25.0$ mag arcsec$^{-2}$ in each band, are
listed in Table 2.  The galaxy has realistic colors, and
its properties agree with recent 2MASS determinations of $B-K$
vs. $M_{K}$ color measurements of nearby spiral galaxies (J. Huchra
2003, private communication).  We also list the stellar mass-to-light
ratios $\Upsilon_{x}$ in each band, for comparison with observed
galaxies.  The stellar mass-to-light ratios in the $g$, $r$, $i$, and
$z$ bands agree well with the properties of observed galaxies in the
Sloan survey \citep{kauffmann03a}.

\begin{figure*}[b]
\centering
\epsfig{file=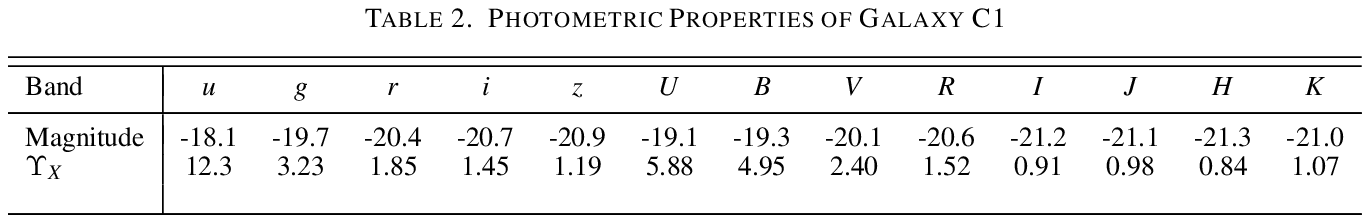,angle=0,keepaspectratio=true}
\end{figure*}

\subsection{Star and metal formation histories}
\label{subsec:sfr}

\begin{figure*}[t]
\centering
\epsfig{file=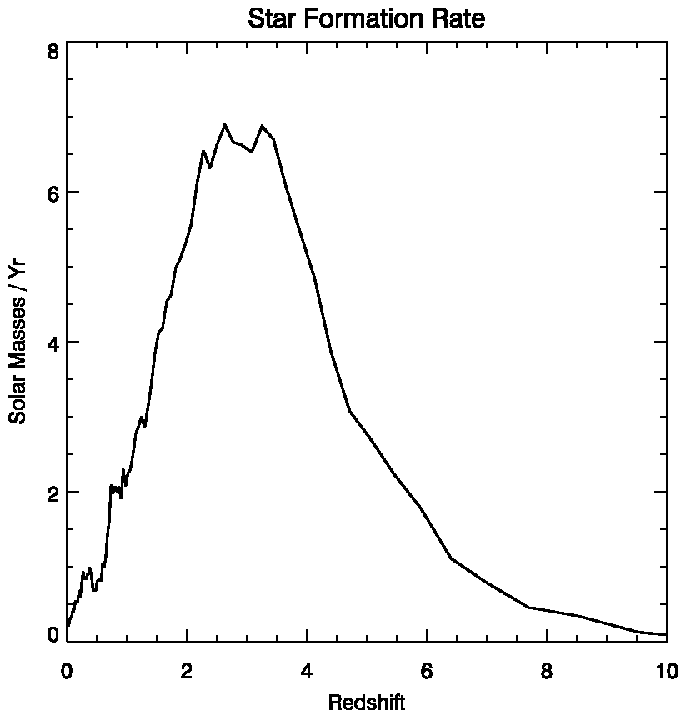,angle=0,totalheight=3.in,keepaspectratio=true}
\epsfig{file=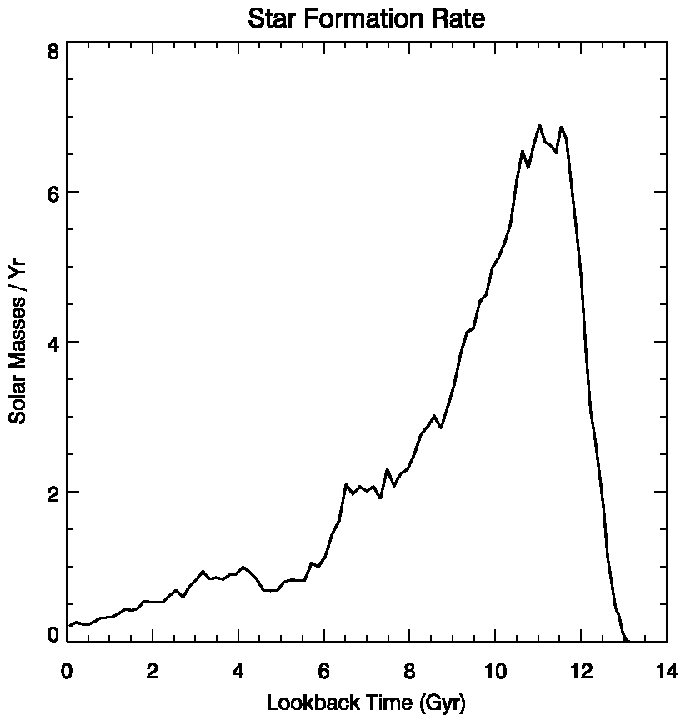,angle=0,totalheight=3.in,keepaspectratio=true}
\vskip -1mm
\figcaption[f6.eps]{
\label{fig:sfz}
Star formation history of the entire simulated galaxy as a function of
redshift (left panel).  The star formation rate in the galaxy
increases from redshift $z=10$ to a peak of $\sim 7$ M$_{\sun}$
yr$^{-1}$ at $z\sim3$.  The right panel shows the galactic star
formation history as a function of lookback time.  Most of the stars
in the galaxy formed earlier than 9 Gyr ago.
}
\vskip -5mm
\end{figure*}

Within our simulation code, we record the formation time of each star
particle, allowing us to infer the detailed star formation history of each object.
Similarly, the metallicity of the star particles can be used to
determine the metallicity evolution of the galaxies.  Figure
\ref{fig:sfz} shows the star formation history of galaxy C1.  As a
function of redshift (left panel), the star formation rate rises below
redshift $z=10$ and peaks at a value $\sim 7$ M$_{\sun}$ yr$^{-1}$ at
redshift $z\sim3$, after which it declines sharply to the present
time.  In the right panel of Figure \ref{fig:sfz}, we also plot the
star formation rate against lookback time.  Viewing the star formation
history in this way illustrates that most of the stars formed early
on, with the majority of stars in the galaxy at $z=0$ being older than
9 billion years. A large fraction of these stars formed in progenitor
halos that later merged to form the galaxy.

\begin{figure*}[t]
\centering
\epsfig{file=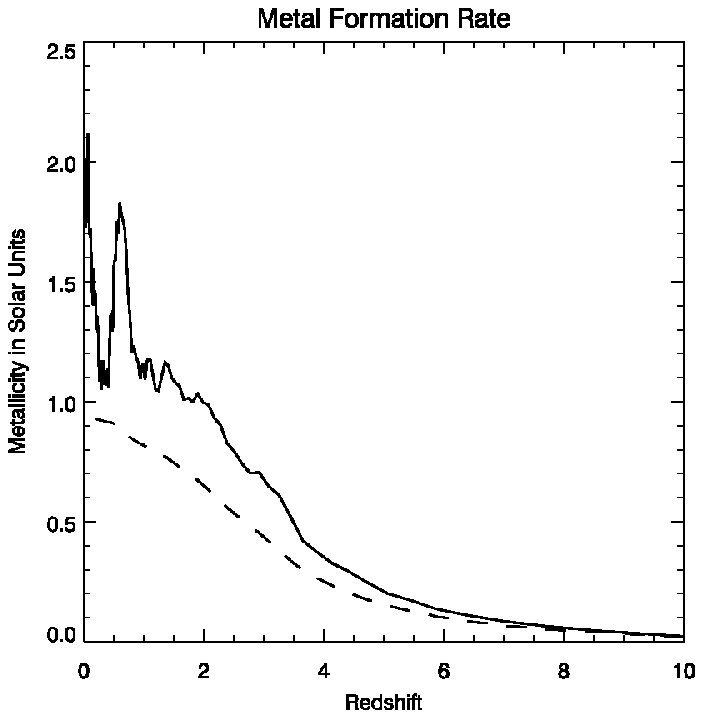,angle=0,totalheight=3.in,keepaspectratio=true}
\epsfig{file=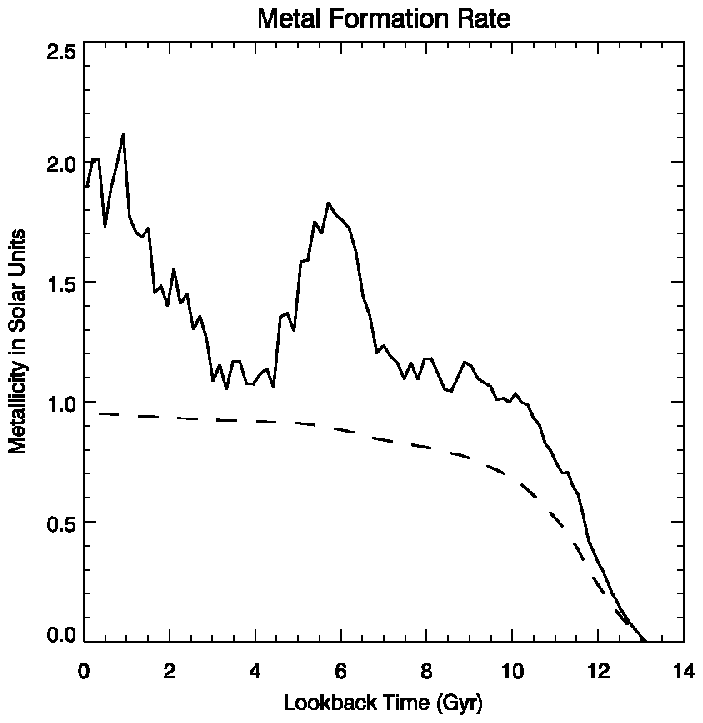,angle=0,totalheight=3.in,keepaspectratio=true}
\vskip -1mm
\figcaption[f7.eps]{
\label{fig:mfz}
Metal formation history of the simulated galaxy. In each panel, the
solid line shows the average mass-weighted metallicity of newly formed
stars over the lifetime of the galaxy.  The dashed line shows the
running mass-weighted average metallicity of all stars within the
galaxy.  The left panel is plotted as a function of redshift and the
right panel is plotted against lookback time.  The large feature at
$z=0.8$ corresponds to a major merger in the assembly history of the
galaxy that significantly altered the metallicity between $z=0.7$ and
$z=0.5$.
}
\vskip -5mm
\end{figure*}

Figure \ref{fig:mfz} illustrates the history of the metal content of
the galaxy.  Each panel shows the average metallicity of newly formed
stars (solid line) over the lifetime of the galaxy and the average
metallicity of all stars formed before each epoch (dashed line).
While the metallicity of newly formed stars varies in a complex
manner, reflecting the details of gas accretion and merging, the
average metallicity of the galaxy evolves smoothly to the current
epoch.  The large feature at $z=0.8$ corresponds to a major merger in
the assembly history of the galaxy that significantly altered the
metallicity between $z=0.7$ and $z=0.5$.  Note that stars formed
before 9 billion years ago have average metallicities of $<0.8
\mathrm{\,\,Z_{\sun}}$.  We hence expect to see an older population of
relatively metal-poor stars, while regions of the galaxy that actively
formed stars even at late times should feature a younger, metal-rich
population. If a disk does not form stars at the same rate everywhere,
we would expect to see stellar age and metallicity gradients between
regions of differing star formation activity.

\subsection{Stellar age and metallicity gradients}
\label{subsec:sag}

\begin{figure*}[t]
\centering
\epsfig{file=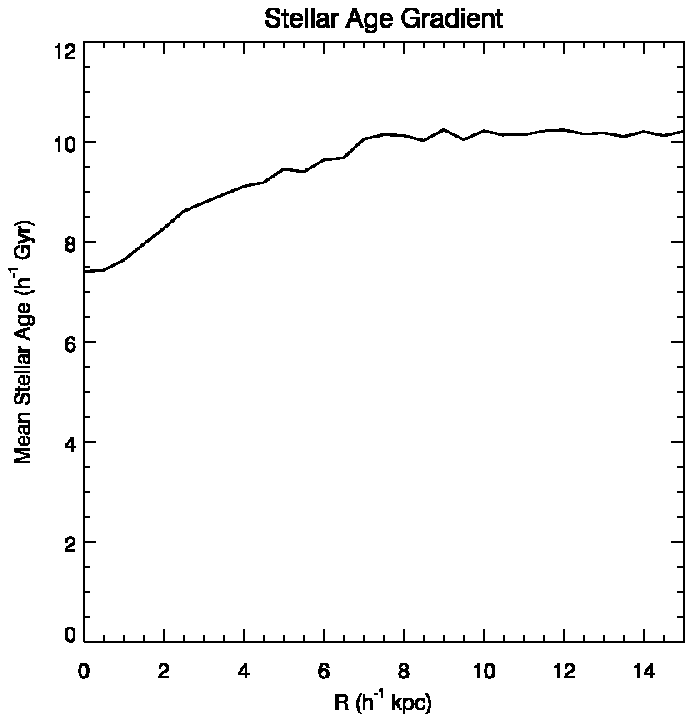,angle=0,totalheight=3.in,keepaspectratio=true}
\epsfig{file=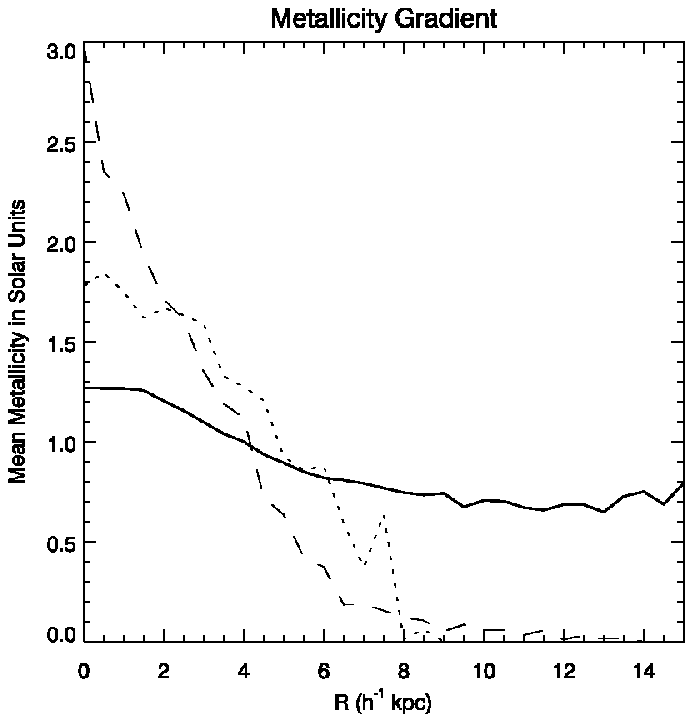,angle=0,totalheight=3.in,keepaspectratio=true}
\vskip -1mm
\figcaption[f8.eps]{
\label{fig:sagsmg}
Stellar age gradient in the disk (left panel), measured by averaging
the ages of star particles in $0.5\,h^{-1}{\rm kpc}$ radial bins.  The
average age of the stars decreases from $\sim 10$ Gyr in the outer
regions of the disk to $\sim 7.5$ Gyr in the center.  The right panel
shows the stellar (solid) and gas (dashed) metallicity gradients in
the disk.  The metallicity of the stars increases from $\sim 0.75$
Z$_{\sun}$ at the edge of the disk to $\sim 1.3$ Z$_{\sun}$ in the
center, while that of the gas in the disk increases strongly from
primordial values in the outer disk to $\sim 2.5$ Z$_{\sun}$ in the
central regions.  Stars forming after $z=0.25$ (dotted line) track the
metallicity gradient of star-forming gas in the disk.
}
\vskip -5mm
\end{figure*}

Combining the formation time and metallicity of each star particle
with information about its position allows us to determine stellar age
and metallicity gradients in the galaxy.  The left panel of Figure
\ref{fig:sagsmg} shows the stellar age gradient in the disk, measured
for all star particles within $15\, h^{-1}$ kpc from the center of the
galaxy.  The average age of the star particles decreases from $\sim
10$ Gyr in the outer regions of the disk to $\sim 7.5$ Gyr in the
center.  Beyond a radius of $\sim 10\,h^{-1}$ kpc, the average age of
the stars is dominated by contributions from the stellar halo, where
stars are significantly older than in the disk.

In the right panel of Figure \ref{fig:sagsmg} we show the metallicity
gradient in the galaxy, plotting the average metallicity of the star
(solid line) and gas (dashed line) components in $0.5 h^{-1}$ kpc
radial bins.  The metallicity gradient of the stars tracks the stellar
age gradient closely, remaining at a roughly constant value of $\sim
0.7 \mathrm{\,\,Z_{\sun}}$ from $r=15 \,h^{-1}$ kpc to $r=8\, h^{-1}$
kpc and then increasing to a central value of Z$=1.3
\mathrm{\,\,Z_{\sun}}$.  Again, the halo stars beyond $10\,h^{-1}{\rm
kpc}$ are older and deficient in metals compared with the younger
stars in the disk.  Interestingly, the gas metallicity gradient is
much stronger than the stellar metallicity gradient, increasing from
close to primordial values at $r = 8\, h^{-1}$ kpc to $\sim
2.6\mathrm{\,\,Z_{\sun}}$ near the center of the disk.  The gas in the
inner regions has been enriched by stars forming at late times (dotted
line, for stars forming after $z=0.25$), whose metallicities have been
enhanced by the long history of star formation in the galaxy.  The
metallicity gradients of the star-forming gas and recently formed
stars in the disk then track each other, again highlighting the strong
metallicity gradient in the gaseous disk.

The active star formation in the disk produces a collection of stars
that are younger and more metal enriched than in the outer regions of
the galaxy.  Hence, the disk is not simply built up inside-out, but
instead partially forms outside-in \citep{slgp03a}.  The more active
star formation at late times in the inner disk makes this region
actually quite young, on average.  In the halo of the galaxy, where
star formation is dormant, stars are significantly older and more
metal poor than in the disk.

\section{Kinematic Properties of Galaxy C1}
\label{sec:disk:kin}

\subsection{Angular momentum}
\label{subsec:jstar}

Perhaps the most important kinematic property of disk galaxies is
their angular momentum, as this quantity determines the sizes of thin,
rotationally supported disks \citep{mmw98a}.  While measuring the
specific angular momentum $j_{\star}$ of each star particle is
straightforward, deciding which particles to use for measuring the
angular momentum content of the disk is less clear, in particular when
comparing with observations.

We wish to compare our simulated galaxy with the $I$-band Tully-Fisher
relation measured by \cite{mfb92a} and \cite{courteau97a} and the
corresponding $j_{\star}-V_{\mathrm{rot}}$ relation \citep[as compiled
by][]{navarro98a}.  The measured rotation velocity $V_{\rm rot}$ used
for these relations is determined observationally at $2.2$ times the
exponential scale-length of the stellar surface brightness, $R_{d}$.
We then take $V_{\rm rot}$ to be
\begin{eqnarray}
V_{\rm rot} &=& \sqrt{\frac{GM(<R_{V})}{R_{V}}},\\
R_{V}   &=& 2.2R_{d,B},
\end{eqnarray}
\noindent
where $R_{d,B}$ is the $B$-band surface brightness scale-length
determined in \S~\ref{subsec:phot}.  In what follows, when we refer to
$V_{\rm rot}$ we always mean the expected rotation velocity
corresponding to the circular velocity produced by all mass within
$2.2 \times R_{d,B}=0.045 R_{200}$ for galaxy C1.

Having determined the rotation velocity $V_{\rm rot}$, we then
consider 3 different approaches for measuring the specific stellar
angular momentum $j_{\star}$, providing slightly different
interpretations of the angular momentum content of the galaxy.  For
all the star particles, we define the total specific stellar angular
momentum $j_{\star,\mathrm{tot}}$ to be a sum over the star particles
of the form
\begin{equation}
j_{\star,\mathrm{tot}} =
{\left|\sum\limits_{i=1}^{N_{\star}}m_{i}\mathbf{r}_{i} \times
    \mathbf{v}_{i}\right|} \;\Bigg/\; {\sum\limits^{N_\star}_{i=1}m_{i}},
\label{eqj1}
\end{equation}
\noindent
where $N_{\star}$ is the total number of star particles within
$R_{200}$, $\mathbf{v}_{i}$ is the velocity with respect to the center
of mass of the galaxy, $\mathbf{r}_{i}$ is the distance from the
center of the galaxy, and $m_{i}$ is the mass of the $i^{\mathrm{th}}$
star particle.

We also consider the specific angular momentum
$j_{\star,\mathrm{R_V}}$ of the stars within the radius $R_{V}$ at
which the rotation velocity is measured.  This is determined by simply
restricting the sums in equation~(\ref{eqj1}) to the stars within a
distance $R_{V}$ of the center of the galaxy.

The final method we employ mimics the observational estimate for the
specific angular momentum content of an exponential disk, viz.
\begin{equation}
j_{\star,\mathrm{obs}} = 2\,R_{d,B}\,V_{\rm rot}.
\end{equation}
\noindent
Unlike the previous two methods, this technique indirectly estimates
the angular momentum content by combining a measurement of the size of
the disk with a typical circular velocity.  While allowing an
inter-comparison of observed galaxies with similar rotation velocities
\citep{abadi03a}, this approach implicitly assumes that the disk has
the rotation curve of an exponential disk \citep{courteau97a}, which
is not true in general.

\begin{figure}[b]
\centering
\epsfig{file=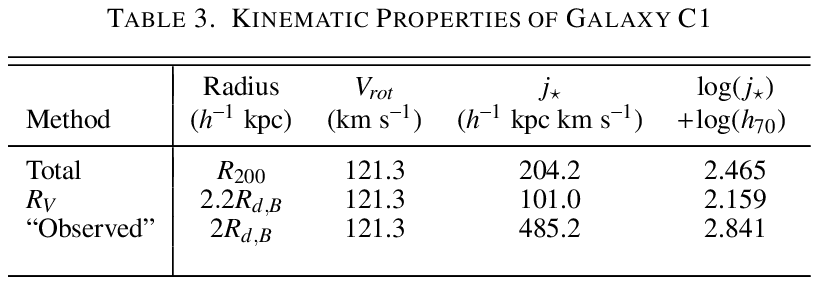,angle=0,keepaspectratio=true}
\end{figure}

We summarize the angular momentum of our simulated galaxy in Table 3,
listing the values for $V_{\rm rot}$ and the
three different measurements of $j_{\star}$ along with their values in
physical units.  With a rotation velocity $V_{\rm rot}=121.3\,{\rm
km\, s^{-1}}$ and total specific stellar angular momentum
$j_{\star}=204.2\, h^{-1}$ kpc km s$^{-1}$, this galaxy agrees
favorably with the observed range of $\log(j_{\star}) \sim 2.4-3.2$
for galaxies with $\log(V_{\rm rot}) \sim 2.1$ \citep{mfb92a,
courteau97a, navarro98a}.  However, the region of the galaxy within
$R_{V}$ has a specific angular momentum lower by a factor of $\sim 2$
with $j_{\star,R_{V}}=101.0 \,h^{-1}\, {\rm kpc\,\, km\,\, s^{-1}}$.
The lower angular momentum of this part of the galaxy results from the
thickness of the stellar disk.  Finally, the agreement between the
value $j_{\star,\mathrm{obs}}=485.2 h^{-1}$ kpc km s$^{-1}$ and the
observed range of $j_{\star}$ demonstrates that the stellar disk has
the proper size when compared with observed galaxies of similar
rotation velocity.

Overall, we find that the total angular momentum content of the galaxy
is appropriate for its rotation velocity, when compared to
observations, but that the thick disk reduces the specific stellar
angular momentum content in the inner regions.  We stress that the
stellar disk of this galaxy should be considered deficient in angular
momentum content and caution that only measuring the total specific
stellar angular momentum content of a simulated galaxy can lead to an
overestimate of the specific stellar angular momentum in the central
regions of the galaxy.

At present, we cannot determine whether the low angular momentum in
the inner regions of the galaxy is purely a physical effect, or is at
least partly a consequence of our relatively poor spatial resolution.
The gravitational softening length in the high-resolution region of
our simulation is $0.65 h^{-1}$ kpc.  For $h=0.7$, this means we
cannot resolve thin stellar disks of galaxies like the Milky Way.  It
is possible that our estimate of the angular momentum of the inner
parts of galaxy C1 is underestimated because of this problem.

\subsection{Rotation curve}

\begin{figure}[t]
\centering
\epsfig{file=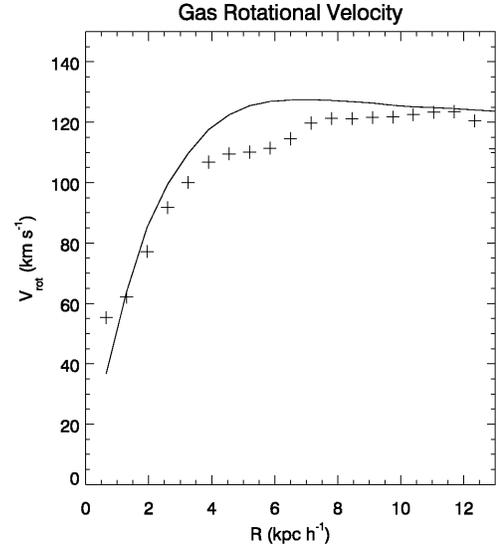,angle=0,totalheight=3.in,keepaspectratio=true}
\vskip -1mm
\figcaption[f9.eps]{
\label{fig:rot_curve}
Rotation velocity of the gas particles within 2 $h^{-1}$ kpc of the
disk plane (crosses), projected onto 0.5 $h^{-1}$ kpc annuli.  We find
that the gas rotation velocity closely traces the total rotation
velocity $V(R) = [GM(<R)/R]^{1/2}$ (solid line), as expected for a
thin disk without a dominant bulge component.
}
\vskip -5mm
\end{figure}

In Figure \ref{fig:rot_curve}, we plot the rotation velocity of gas
particles from within 2 $h^{-1}$ kpc of the disk plane (crosses),
projected onto 0.5 $h^{-1}$ kpc wide annuli.  We find that the
rotation velocity of the gas closely traces the total rotation
velocity $V(R) = [GM(<R)/R]^{1/2}$ (solid line), as expected for a
thin disk without a dominant bulge component.  If the central regions
of the galaxy contained a kinematically important bulge, the gas
rotation curve would either rise steeply or have a central maximum and
decline outwards.  We interpret the slowly rising rotation velocity of
the gas as further evidence that galaxy C1 consists of a rotationally
supported exponential disk without a significant bulge component.

\begin{figure*}
\centering
\epsfig{file=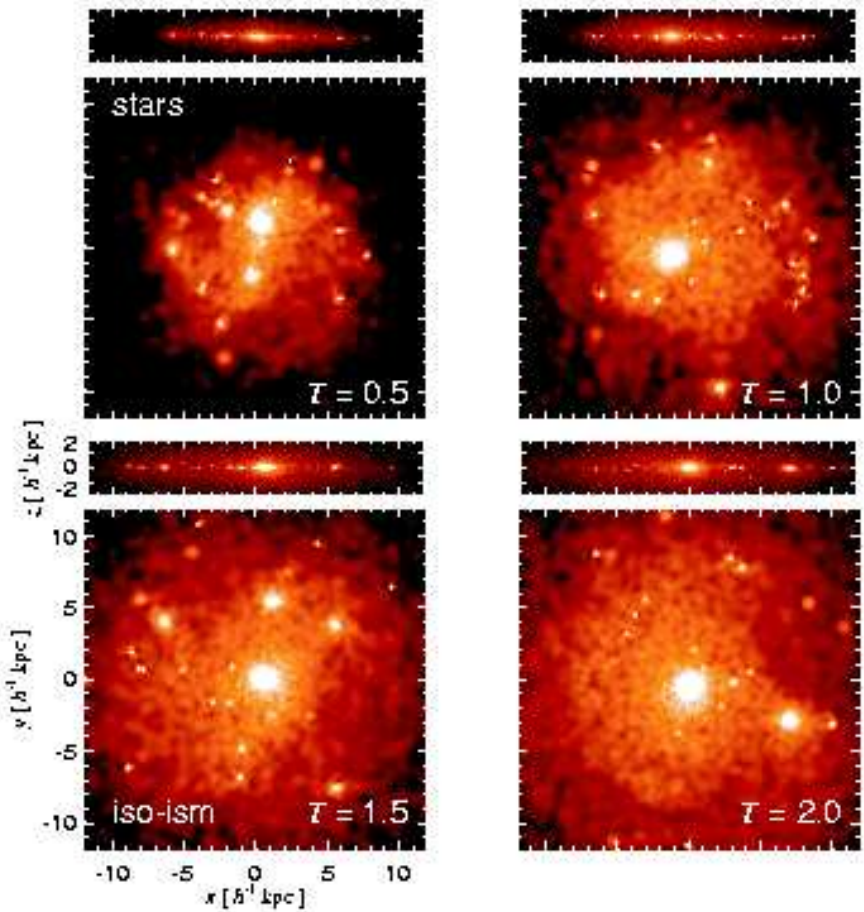,angle=0,totalheight=5.in,keepaspectratio=true}
\vskip -1mm
\figcaption[f10.eps]{
\label{fig:d4:iso-ism}
Stellar mass distribution, adaptively smoothed and seen edge- and
face-on, for the ``iso-ism'' model at (clockwise from upper left)
$T=0.5$ Gyr, $T=1.0$ Gyr, $T=1.5$ Gyr, and $T=2.0$ Gyr.  As the disk
develops, gas quickly cools to roughly $10^{4}$ K and forms stars.
The $10^{4}$ K gas is Toomre unstable and fragments into dense clumps,
which in turn efficiently form stars, leading to the clumpy stellar
mass distribution in the disk.  As the galaxy ages, the clumps lose
angular momentum through dynamical friction, coalesce, and fall into
the center of the galaxy.
}
\vskip -5mm
\end{figure*}

\begin{figure*}
\centering
\epsfig{file=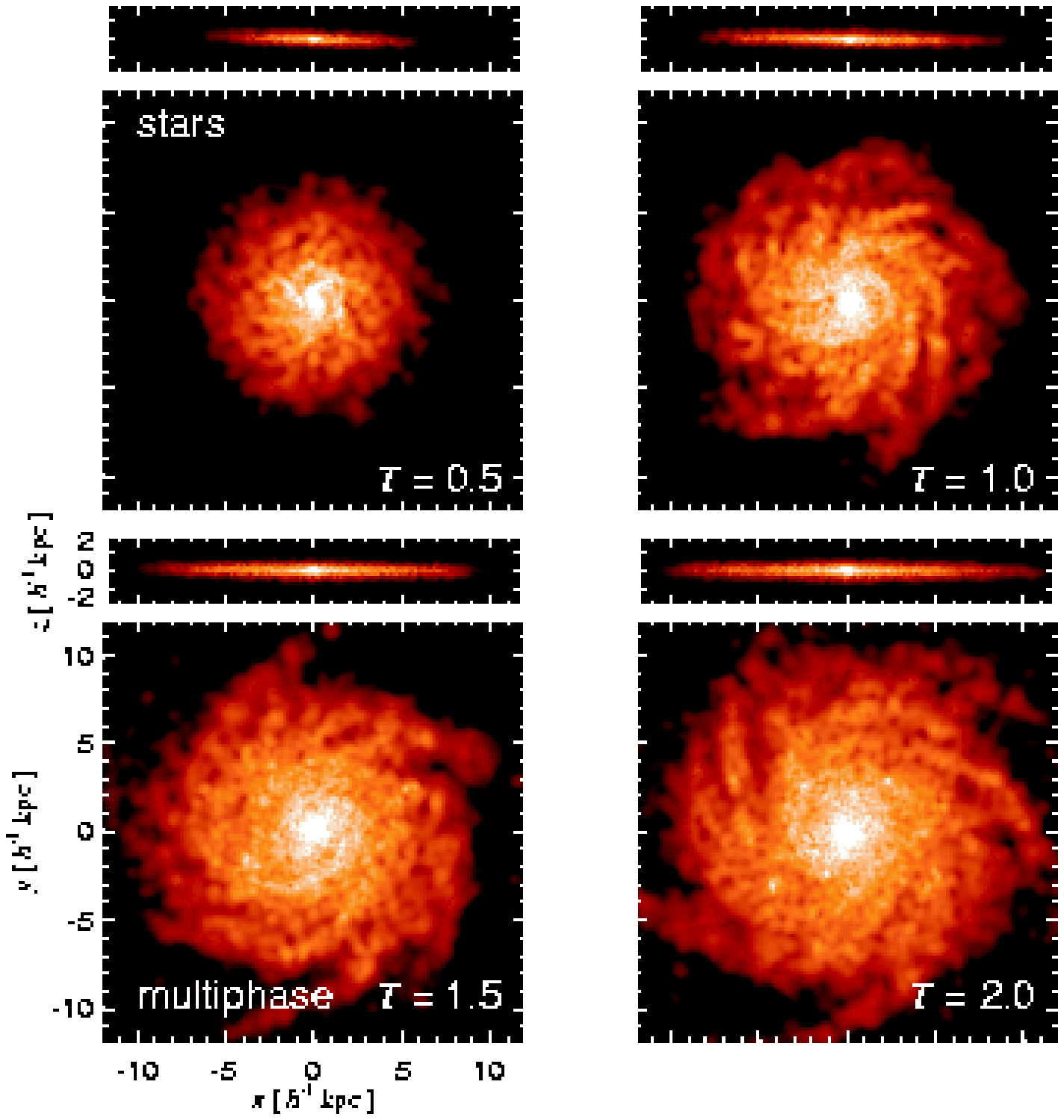,angle=0,totalheight=5.in,keepaspectratio=true}
\vskip -1mm
\figcaption[f11.eps]{
\label{fig:d4:multiphase-ism}
Stellar mass distribution, adaptively smoothed and seen edge- and
face-on, for the ``multiphase-ism'' model at (clockwise from upper
left) $T=0.5$ Gyr, $T=1.0$ Gyr, $T=1.5$ Gyr, and $T=2.0$ Gyr.  In the
``multiphase-ism'' run, the gas cools into a thin disk which begins
forming stars at a rate regulated by supernova feedback.  While the
gas is comprised of clouds cool enough to form stars, the
contributions from the hot surrounding medium to the effective
equation of state of the gas provides pressure support to the ISM to
prevent Toomre instability.  The end result is a large disk galaxy
with only a modest bulge component.
}
\vskip -5mm
\end{figure*}

\subsection{$I$-band Tully-Fisher relation}
\label{subsec:itf}

From the photometric modeling described in \S~\ref{subsec:phot} and
the kinematic measurements in \S~\ref{subsec:jstar}, it is possible to
compare galaxy C1 with the $I$-band Tully-Fisher relation as
determined by \cite{giovanelli97a}, \cite{hm92a}, and \cite{mfb92a}.
For our measured $I$-band magnitude of $I=-21.2$ and rotation velocity
$V_{\rm rot}=121.3$ km s$^{-1}$, we find our galaxy to be in good
agreement with the average observed $I$-band magnitude of $I \sim
-20.8$ for galaxies with $\log(V_{\rm rot})\sim 2.1$.

\section{Isolated Halo Simulations}
\label{sec:sim:halo:intro}

The formation of a well-defined disk galaxy without a dominant bulge
can be viewed as a significant success towards obtaining realistic
galaxies in cosmological simulations.  It is therefore of particular
interest to understand which aspects of our modeling are mainly
responsible for this outcome.

A novel feature of our approach lies in our use of a subresolution
model for the multiphase structure of the ISM, together with an
effective equation of state that this model implies for the dense
star-forming gas.  As can be seen in Figure \ref{fig:eos}, when the
gas is sufficiently dense to be thermally unstable and subject to star
formation, the equation of state departs from that of an isothermal
gas, and becomes stiffer with increasing density.  This will affect
the bulk hydrodynamics of the gas, through the pressure gradient term
in the Euler equation.  Consequently, dense gaseous disks will
experience additional pressure support from supernova feedback, in a
manner that is different from previous attempts to model galaxy
formation.

In order to examine the importance of this effect further, we use a
set of simulations to consider disks forming in isolated halos
\citep[e.g.][]{sh03a}, where we vary the description of star-forming
gas and associated feedback effects.  These results show that the
physics of dense, star-forming gas has a substantial impact on the
dynamics of disks, indicating that the model of the ISM plays a
crucial role in our successful formation of a disk galaxy in a
cosmological context.

\subsection{Initial conditions and simulation set}

\label{sec:sim:halo}

In order to isolate the dependence of disk formation on the physics of
the ISM, we investigate the dynamics of gas in static spherically
symmetric NFW \citep{nfw95a} dark matter halos of mass $10^{12}h^{-1}$
$\Msun$ and concentration $c_{\mathrm{NFW}}=20$.  We place $40,000$
gas particles that initially trace the dark matter in these halos and
set the initial temperature profile such that the gas is in
hydrostatic equilibrium when evolved with non-radiative hydrodynamics.
We add angular momentum to the gas, corresponding to a relatively high
spin parameter of $\lambda =
J|E|^{1/2}/(GM_{\mathrm{vir}}^{5/2})=0.1$.  We then evolve the gas
forward in time including radiative cooling.  Under these conditions,
we expect the gas to collapse smoothly into a thin, rotationally
supported disk.  We do not intend for this to be a realistic model of
disk formation in a hierarchical universe, but the idealized nature of
the simulations described below makes it possible to examine the
consequences of varying the gas physics in detail.

This set-up provides initial conditions for a suite of simulations,
where we consider three different descriptions of the ISM.  The first
model, which we will refer to as ``justcool,'' consists of simply
allowing the gas to cool with primordial abundances without allowing
for any star formation or feedback.  The ``justcool'' scenario serves
as an extreme reference case where star formation and feedback cannot
affect the gas dynamics.  Since the dense gas that settles into a disk
can cool efficiently, its temperature will remain close to $10^{4}$ K,
effectively obeying an isothermal equation of state.

Our second description of the ISM, which we refer to as ``iso-ism,''
includes gas cooling and star formation, according to
equation~(\ref{eq:sfr}), but no feedback effects.  Gas in the disk
again obeys a nearly isothermal equation of state in this case, but
compared to our ``justcool'' model, dense gas is converted into
collisionless star particles. For ease of comparison, we use the same
star formation rate for gas of a given density as in our multiphase
model.

Finally, our third description of the ISM is the \cite{sh03a}
multiphase ISM model, which includes radiative cooling, star
formation, and supernova feedback in the form of thermal energy input
and cloud evaporation, as described earlier.  We refer to this model
as the ``multiphase-ism'' model.  The primary difference between the
``multiphase-ism'' and ``iso-ism'' scenarios is that in the multiphase
model feedback from supernovae modifies the effective equation of
state of star-forming gas, providing increased pressure support, as
indicated by Figure \ref{fig:eos}.  The stiffer equation of state
helps to regulate star formation so that it occurs at a rate
consistent with observations of isolated disk galaxies
\citep{kennicutt89a,kennicutt98a,mk01a} once the free parameter
$t_0^\star$ of the model is properly adjusted
\citep[see][]{sh03a,sh03b}.

For each ISM model, we ran a number of simulations, varying the
numerical integration parameters to test the robustness of the
results.  In particular, we varied the number of SPH neighbors and the
value of the time-step taken by the code. None of these choices had a
significant effect on the outcome.  Therefore, our results appear to
have converged with respect to time integration.  When the number of
SPH neighbors is varied, instabilities in the disks tend to occur at
slightly different times and locations, but the qualitative behavior
of the simulations remains unchanged.

\subsection{Comparison of disks formed in isolated halos}
\label{sec:halos}

The time evolution of simulations with our three ISM models reveals
interesting differences.  In Figure~\ref{fig:d4:iso-ism}, we show the
stellar mass distribution, adaptively smoothed and seen edge- and
face-on, for the ``iso-ism'' model at (clockwise from upper left)
$T=0.5$ Gyr, $T=1.0$ Gyr, $T=1.5$ Gyr, and $T=2.0$ Gyr.  Gas in the
inner parts of the halo quickly cools to roughly $10^{4}$ K, settles
into a thin, rotationally supported disk, and forms stars.  The
$10^{4}$ K gas in the disk is highly compressible, obeying a nearly
isothermal equation of state.  Consequently, the self-gravitating
gaseous disk becomes gravitationally unstable as soon as it reaches a
moderate surface density.  The \cite{toomre64a} instability quickly
causes a fragmentation of the smooth gas disk into clumps, which
collapse further under their own self-gravity and form stars on a
short time-scale.  As the galaxy ages, the stellar clumps coalesce
through dynamical friction into ever larger lumps, destroying the
disk.  By $T=2\,{\rm Gyr}$, the stars are mainly confined in two large
blobs, with numerous smaller clumps orbiting the center.

The evolution of the galaxy in the ``iso-ism'' model contrasts sharply
with that in the corresponding ``multiphase-ism'' run.  We illustrate
this in in Figure~\ref{fig:d4:multiphase-ism}.  Again, the gas cools
into a thin, star-forming disk. However, supernova feedback regulates
the rate of star formation and the dynamics of the dense star-forming
gas.  Owing to feedback, the ISM is pressurized, stabilizing the gas
against the Toomre instability.  Consequently, a smooth distribution
of gas is maintained in the plane of the disk and stars form steadily,
increasing the size of the disk from $6\, h^{-1}\, {\rm kpc}$ at
$T=0.5$ Gyr to $>12\, h^{-1}\,{\rm kpc}$ at $T=2.0\,{\rm Gyr}$.  Since
the gas remains stable and does not fragment into clumps, the disk
largely avoids angular momentum loss and the simulation yields a large
disk galaxy with only a modest bulge.

\begin{figure*}[t]
\centering
\epsfig{file=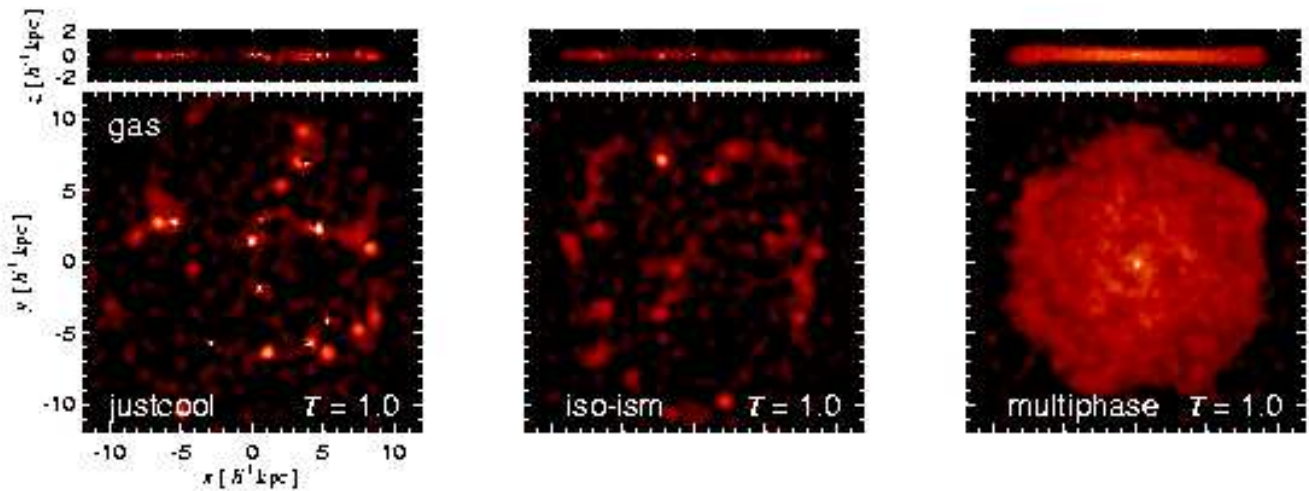,angle=0,totalheight=3.in,keepaspectratio=true}
\vskip -1mm
\figcaption[f12.eps]{
\label{fig:3ismz0}
Comparison of the gas mass distribution in the ``justcool'' (left),
``iso-ism'' (center), and ``multiphase-ism'' (right) simulations at
$T=1.0$ Gyr.  Both the ``justcool'' and ``iso-ism'' models produce gas
disks fragmented by catastrophic Toomre instability.  By contrast, the
``multiphase-ism'' model yields a stable disk supported by additional
pressure supplied by the \cite{sh03a} model for the ISM.
}
\vskip -5mm
\end{figure*}
In Figure \ref{fig:3ismz0}, we compare the gas mass distributions in
the ``justcool'' (left), ``iso-ism'' (center), and ``multiphase-ism''
(right) simulations at $T=1.0$ Gyr.  Both the ``justcool'' and
``iso-ism'' models produce clumpy gas disks that fragmented by the
Toomre instability.  In fact, the time evolution of the ``justcool''
and ``iso-ism'' models is very similar, except that in the former case
the instabilities occur earlier, showing that star formation can have
a stabilizing effect by reducing the gas surface density.  Note,
however, that only the ``multiphase-ism'' model produces a gaseous
disk that exhibits long-term stability.

In summary, our simulations of the formation of disks in isolated
halos clearly show that the modeling of the ISM has a strong effect on
disk stability.  In particular, an isothermal equation of state for
star-forming gas produces gaseous disks that are violently unstable.
Under these conditions it would be difficult if not impossible to form
large, extended disk galaxies unless the gas surface density is always
kept extremely low, for example by adopting an unrealistically short
star formation time-scale that would conflict with the observed
Kennicutt law.

In contrast, the multiphase model for the ISM yields disk galaxies
that are more stable.  The stiffer equation of state produces disks
that are more resilient against Toomre instability than in models with
an effectively isothermal ISM.  Fragmentation is avoided, enabling the
steady formation of large stellar disks even when the disk is gas-rich
and has a high gaseous surface density.  Therefore, we suggest that
the behavior of the equation of state in our multiphase description of
the ISM is the primary reason why our cosmological simulation can
produce disk galaxies with negligible bulge components.  This
conclusion should be insensitive to the detailed features of our
multiphase model of the ISM, provided that the actual equation of
state for star-forming gas behaves similarly to the one proposed by
\citet{sh03a}.

\section{Conclusions}
\label{sec:conclusions}

We have performed a high-resolution cosmological simulation that
included gas dynamics, radiative heating and cooling, and star
formation and supernova feedback, modeled in the framework of a
subresolution description of a multiphase ISM.  Inspecting the ten
largest halos, we identified a well-resolved disk galaxy at redshift
$z=0$.  The disk galaxy has realistic photometric properties,
including a reasonable central surface brightness and exhibits good
agreement with the $I$-band Tully-Fisher relation.  Most important,
this object is the first published example of a disk galaxy from a
simulation that features a simple exponential surface brightness
profile without having a large bulge component.  Our work demonstrates
that extended disk galaxies can form in a $\Lambda$CDM cosmology
within full hydrodynamic simulations, an important step forward in
understanding the origin of disk galaxies.

While the total angular momentum of the stellar component of the disk
galaxy is appropriate for observed galaxies with similar rotation
velocities, the inner regions of the galaxy have low specific stellar
angular momentum.  It is not clear whether this is a problem with the
physics, e.g.~excessive angular momentum loss by the gas, or is at
least partly an artifact of our inability to completely resolve the
thin stellar disk.  The disk galaxy in our simulation does not have an
obvious problem with its radial size, but statistical studies with
samples of simulated disk galaxies will be needed to settle this
question.  However, we find the success achieved here encouraging and
suggest that the problem with disk size is alleviated by employing
equations of state for star-forming gas that are stiffer than
isothermal.

Using a suite of isolated galaxy formation simulations, we have
explicitly demonstrated the consequences our treatment for the
ISM has on the formation of disk galaxies.  Simulations with an
effectively isothermal ISM yield gravitationally unstable disks and
fail to produce large, smooth stellar disks.  In contrast, the
multiphase model supplies enough pressure support to the star-forming
gas to prevent catastrophic Toomre instability, allowing stable
galactic disks to form.  We have also verified that these conclusions
do not depend on details of our numerical integration.

Our work differs in a number of respects from earlier numerical
attempts to study disk galaxy formation.  We have used a novel
formulation of smoothed particle hydrodynamics that, by construction,
conserves both energy and entropy simultaneously, even when smoothing
lengths vary in response to changes in density \citep{sh02a}. While we
consider it unlikely that previous numerical work on galaxy formation
with SPH was strongly affected by inaccuracies in e.g.~entropy
conservation, it is clearly preferable to use a fully conservative
scheme that mitigates against any numerical ``overcooling.''

More important, we have formulated the consequences of feedback using
a subresolution model for star formation in a multiphase ISM.  In this
scheme, the local supernova feedback modifies the effective equation
of state for star-forming gas, leading to an efficient self-regulation
of star formation and a stabilizing effect on highly overdense gas.
This aspect of our modeling feeds directly into the dynamics of
forming disks by modifying the pressure gradient term in the
hydrodynamic equations of motion.

Our formulation of the ISM physics has a number of advantages over
earlier treatments of star formation and feedback.  Other workers have implemented techniques to
allow feedback to have a significant
dynamical impact by, e.g.~depositing thermal energy into gas
surrounding star-forming regions and artificially delaying the
radiative loss of this energy by the gas for an ad hoc time interval.
While similar in spirit to our method, these approaches 
make the comparison of results from different calculations challenging
because the
consequences of feedback are difficult to summarize in a concise,
quantitative form.  When expressed in terms of an effective equation
of state, as done here, the dynamical implications of feedback become
easier to interpret.  A discussion in terms of the effective equation
of state is also relatively insensitive
to the detailed physics responsible for pressurizing the gas, and
enables one to relate the description of feedback to
observations of the ISM.

In order to demonstrate the sensitivity of galaxy formation to the
physics of star-forming gas, we employed a simple multiphase model of
the ISM \citep{sh03a}, where the gas is imagined to be thermally
unstable at sufficiently high densities and exists in two distinct
phases, in pressure equilibrium.  In detail, it is not plausible that
this is a true picture of the ISM of disk galaxies, because this model
ignores, e.g.~turbulent motion, magnetic fields, and cosmic rays.
However, we believe that the basic points of our modeling do not
depend on these complications.  The evolution of the different models
shown in \S~\ref{sec:sim:halo:intro} depends mainly on the effective
equation of state for the star-forming gas on scales which
characterize the Toomre instability.  According to the Toomre
criterion, a thin sheet of gas in differential rotation will be
gravitationally unstable if
\begin{equation}
Q \, \equiv \, {{c_s \kappa}\over {\pi G \Sigma}} \, < \, 1 \, ,
\end{equation}
where $c_s$ is the sound speed, $\kappa$ is the epicyclic frequency,
and $\Sigma$ is the surface density.  In terms of an effective
temperature of the gas,
\begin{equation}
Q \, \approx \, 1.1 \, T_5^{1/2} \, \kappa_{35} \, \Sigma_{75}^{-1}
\, ,
\end{equation}
where $\kappa_{35} \equiv \kappa/ (35\,{\rm km \, sec^{-1}kpc^{-1}})$
and $\Sigma_{75} \equiv \Sigma / (75\,{\rm M}_\odot\,{\rm pc}^{-2})$
are values characteristic of the Milky Way in the solar neighborhood,
and $T_5 \equiv T/(10^5\,{\rm K})$.  If the gas obeyed an isothermal
equation of state with $T_5 = 0.1$, then $Q\sim 0.3$, and the disk
would be violently unstable, as for example in the case of our
``iso-ism'' simulations described in \S~6.  With our equation of
state, however, the effective temperature at densities characteristic
of a forming galaxy would be $T_5 \sim 1$ \citep[see Fig.~1
of][]{sh03a}, so $Q>1$ and the disk would be stable, explaining why
our ``multiphase-ism'' models in \S~6 produce realistic disks.
Moreover, for thin disks of gas, the fastest growing unstable mode has
a wavelength which is a significant fraction of the disk size
\citep[e.g.][]{bt87} and, therefore, the growth of the instability is
sensitive only to the macroscopic equation of state for the ISM.
Whether or not the effective equation of state we have adopted
(i.e.~Figure~\ref{fig:eos}) for the ISM is appropriate in detail is
uncertain, but this question is largely independent of the exact
properties of star-forming gas.  In particular, we expect our results
to generalize to other models for the ISM with similar effective
equations of state.

Having identified one exponential disk galaxy in our simulation,
future work is needed to extend our analysis to other galaxies.  For
example, a good statistical sample of simulated disks is needed before
we can characterize the distribution of disk sizes and their
morphological types, as well as the photometric and kinematic
properties of the galaxies.  This work will require larger simulations
of substantial dynamic range in order to obtain adequate numbers of
simulated disk galaxies, but we find the initial successes described
here and recently by other authors highly encouraging.  A comparison
with the wealth of observational data on the structural parameters of
galaxies in the local universe will then show whether hydrodynamical
simulations are finally producing realistic disk galaxies.  \\ \\

This work was supported in part by NSF grants ACI 96-19019, AST
98-02568, AST 99-00877, AST 00-71019, and AST-0206299, and NASA ATP
grants NAG5-12140, NAG5-13292, and NAG5-13381.  NY acknowledges
support from JSPS Special Research Fellowship (02674).  The
simulations were performed at the Center for Parallel Astrophysical
Computing at the Harvard-Smithsonian Center for Astrophysics.

\end{document}